\documentclass[submission,copyright,creativecommons]{eptcs}
\providecommand{\event}{AiSoS 2013} 
\usepackage{breakurl}             
\usepackage{xcolor,colortbl}
\usepackage{graphicx}
\usepackage[normalem]{ulem}

\title{SoS contract verification using statistical model checking}
\author{ Alessandro {\sc Mignogna}, \\Leonardo {\sc Mangeruca}
\institute{
ALES S.r.l., \email{First.Last@ales.eu.com}
}
\and
Beno\^{i}t {\sc Boyer}, Axel {\sc Legay}
\institute{INRIA - Rennes Bretagne Atlantique\\ Rennes, France}
\email{First.Last@inria.fr}
\and
Alexandre {\sc Arnold}
\institute{EADS}
\email{alexandre.arnold@eads.net}
}

\def\titlerunning{SoS contract verification using statistical model checking}
\def\authorrunning{A. Mignogna, L. Mangeruca, A. Legay, B. Boyer, A. Arnold}
\begin{document}
\maketitle

\begin{abstract}
 Exhaustive formal verification for systems of systems (SoS) is impractical and cannot be applied on a large scale. In this paper we propose to use statistical model checking for efficient verification of SoS. We address three relevant aspects for systems of systems: 1) the model of the SoS, which includes stochastic aspects; 2) the formalization of the SoS requirements in the form of contracts; 3) the tool-chain to support statistical model checking for SoS. We adapt the SMC technique for application to heterogeneous SoS. We extend the UPDM/SysML specification language to express the SoS requirements that the implemented strategies over the SoS must satisfy. The requirements are specified with a new contract language specifically designed for SoS, targeting a high-level English-pattern language, but relying on an accurate semantics given by the standard temporal logics. The contracts are verified against the UPDM/SysML specification using the Statistical Model Checker (SMC) PLASMA combined with the simulation engine DESYRE, which integrates heterogeneous behavioral models through the functional mock-up interface (FMI) standard. The tool-chain allows computing an estimation of the satisfiability of the contracts by the SoS. The results help the system architect to trade-off different solutions to guide the evolution of the SoS.
\end{abstract}


\section{Introduction}
As systems and their interactions grew in complexity, it became more and more important to recognize new entities that were given the suggestive name of systems of systems. With the identification of systems of systems as new entities that were somewhat different from systems, there came the necessity of introducing a definition for these new entities. It became immediately apparent that a clear boundary between system and system of systems cannot be formally identified and the character of systems of systems, called SoS-ness, can only qualitatively be outlined. A solution to this paradox was proposed in \cite{Maier1998} by introducing a series of characteristics that help highlighting the differences. These characteristics are not binary and allow for different shades of SoS-ness. Maier’s characteristics can be described in terms of five categories: emergent behavior, geographic distribution, evolutionary development, operational independence, managerial independence. The Open Group Architecture Forum \cite{opengroup} later acknowledged these characteristics in their description:
{\it Systems of Systems should be distinguished from large but monolithic systems by the independence of their elements, their evolutionary nature, emergent behaviours, and a geographic extent that limits the interaction of their elements to information exchange.}

The characteristics of the SoS lead to new challenges compared to the systems engineering paradigm. These new challenges are responsible for a higher level of complexity with respect to traditional systems engineering, due to multiple complexity-raising factors, such as very large number of components, huge state space, inherent stochastic behavior, intrinsic structural and functional evolution.

The arduous consequences of these SoS traits include the strenuous efforts required for the verification of SoS behaviors. Classical formal verification approaches rely on the implicit exploration of the reachable state space, which is highly impractical for systems of systems. Although, abstraction techniques may mitigate the complexity associated with the verification of SoS, they require substantial modeling and model validation effort. 

In this paper, we describe the DANSE \cite{project-DANSE} approach to the verification of SoS. The approach is based on statistical model checking (SMC) \cite{LegaySMCOverview} and is limited to performance verification to reduce the required accuracy of the underlying statistical inference mechanisms, in order to keep the technique scalable with the ever increasing complexity of SoS.

\section{SoS Modeling}

In the scope of the DANSE project, we selected UPDM/SysML \cite{UPDM} as the reference language for SoS modeling. The main rationale lies in the fact that UPDM originated from and DoDAF \cite{DoDAF} and MODAF \cite{MODAF} and that UPDM is an extensible language.

\subsection{UPDM overview}
UPDM is a unified Profile for DoDAF (Department of Defence Architecture Framework) and MODAF (Ministry of Defence Architectural Framework). It supports the capabilities to model architectures of complex systems, System of Systems, and service oriented architectures.

\begin{figure}[h]
  \centering
  \includegraphics{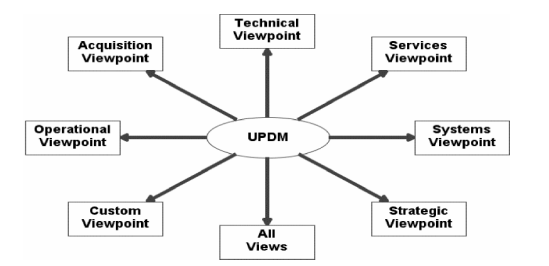}
  \caption{UDPM viewpoints}
  \label{fig:UDPMviewpoints}
\end{figure}

In UPDM model elements are organized in various viewpoints and views. The views and viewpoints which are defined in UPDM are depicted in Fig.~\ref{fig:UDPMviewpoints}. All these views and viewpoints consist of further sub-views such that the contained models are more focussed on certain criteria of the design.
In the following we will give a brief description to all views and viewpoints.  

\begin{itemize}

\item The {\bf Acquisition and Project Views} (AcV/PV) describe project details, e.g. dependencies between projects. AcV/PV contains the following sub-views: AcV-1/PV-1 (organizational perspective on projects), AcV-2/PV-2 (timeline perspective on projects), and PV-3 (projects realizing capabilities);

\item The {\bf Operational View} (OV) is about real-world activities and “answer the “who,” “what,” “when,”, “where,” “why,” and “how” of a mission”. Contained sub-views: OV-1 (mission or scenario description, list of operational elements), OV-2 (operational node relation description), OV-3 (operational information exchange matrix), OV-4 Actual (relations among resources), OV-4 Typical (organizational structures and interactions), OV-5 (operational activity model), OV-6a (operational rule model), OV-6b (operational state transition description), OV-6c (operational event-trace description), and OV-7 (information models on operational architecture);

\item The {\bf All Views} (AVs) provide an overview description of the considered architecture. Also scope, ownership, and timeframe are represented here. {\it “The AVs include a dictionary of the terms used in the construction of the architecture”}. AV-1 (overview information), and AV-2 (representation of all elements of architecture as standalone structures);

\item The {\bf Strategic Viewpoint} (StV) helps to manage the capability management by providing an overall Enterprise Architecture assessment of the corresponding capabilities and their relationships. In this view, capabilities are introduced, their integration is described, and the realignment or removal are modelled. Contained sub-views: CV-7 (mapping of capabilities and services), StV-1 (strategic context for enterprises), StV-2 (capability taxonomies), StV-3 (capability phasing), StV-4 (dependencies between capabilities), StV-5 (fulfilment of capability requirements), and StV-6 (mapping of capabilities and operational activities);

\item The {\bf Systems Viewpoint} (SV) describes realizations of architectures such as resource interaction specifications (SV-1/SvcV-1) or defines specifications on functional and non-functional aspects. The models within this viewpoint {\it “represent alternate realizations in terms of equipment capability of the operational capabilities expressed through models in the Operational Viewpoint and in the User Requirements''}. Contained sub-views: SV-1/SvcV-1 (resource interaction specification), SV-2/SvcV-2 (systems communication description), SV-3/SvcV-3a (resource interaction matrix), SV-4/SvcV-4 (functionality description), SV-5/SvcV-5 (implementation of operational activities), SV-6/SvcV-6 (system data exchange matrix), SV-7/SvcV-7 (resource performance parameters matrix), SV-8/SvcV-8 (change of capability configurations), SV-9/SvcV-9 (technology forecast), SV-10a/SvcV-10a (functional, non-functional specifications), SV-10b/SvcV-10b (resource response description), SV-10c/SvcV-10c (description of interactions between resources), SV-11/DIV-3 (definition of structure of system data), and SV-12 (service provision);

\item The {\bf Service-Orientated View} (SOV) is a description of services offered by constituent systems, which are needed to support the operational domain, which are described in the OV. Contained sub-views: SOV-1 (service hierarchy and taxonomy), SOV-2 (service specification), SOV-3 (service mapping view), SOV-4a (service constraints view), SOV-4b (service state model), SOV-4c (service interaction specification), and SOV-5 (service functionality);

\item The {\bf Technical Viewpoint} consists of elements describing standards, rules, notations, and conventions “that apply to the implementation of the system architecture”.  Contained sub-views: TV-1 (technical standards), and TV-2 (technology standard changes).
\end{itemize}

For more detailed information on the UPDM specification please refer to [5]. 

\subsection{Stochastic modeling in UPDM/SysML}
Our proposal is based on a set of attribute stereotypes that can be applied to any block attribute. This idea is close to the suggestion of the non-normative distribution extensions made in appendix of the SysML 1.3 specification \cite{SysML}, but adds the possibility to regenerate a distribution-based random value whenever needed (and not only at initialization). Being able to regenerate a random number based on a same distribution allows expressing random processes to model real-world behaviours, such as the time needed for a human to perform a task, which is repeated over time. Our proposal is driven by the following requirements:

\begin{itemize}
\item Simple enough for the end-user (the SoS architect);
\item Easily readable and exploitable by the run-time technologies;
\item Consistent with well-known mathematical theory;
\item Flexible to accommodate user-defined distributions;
\item Expressive to allow modeling Markov chains and processes.
\end{itemize}

In order to include stochastic aspects into the UPDM/SysML model, the following
stereotypes are proposed for the DANSE profile:

\begin{itemize}
\item Uniform distribution $\rightarrow$ min and max properties;
\item Normal distribution $\rightarrow$ mean and standard deviation properties;
\item Custom distribution $\rightarrow$ custom observation function property (user-defined).
\end{itemize}

These three different kinds of generic random variables are a compromise between:

\begin{itemize}
\item Usability $\rightarrow$ pick one variable kind from a set of predefined ones (library);
\item Flexibility $\rightarrow$ the user can specify custom distributions;
\item Expressiveness $\rightarrow$ the above properties can be defined as ‘String’, the random
variables could support expressions which refer to the model, which especially en-
ables defining causal dependencies to the state of the model (e.g. take time into ac-
count).
\end{itemize}

\begin{figure}[bth]
  \centering
  \includegraphics{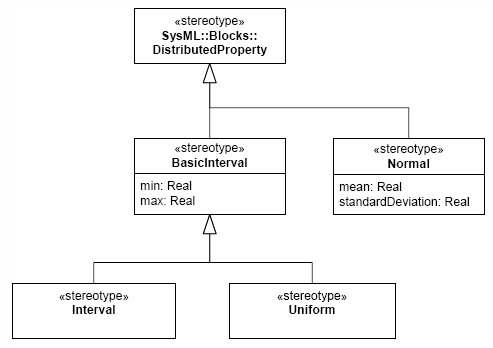}
  \caption{Stereotypes to represent stochastic models}
  \label{fig:stereotypes}
\end{figure}

Note that this approach also covers the ability to express the probability that one tran-
sition is taken among a finite subset of transitions in the behavioral model. To achieve
this objective it is sufficient to introduce a random variable and use its value in transi-
tion guards to select among the different transitions in the subset.
Stochastic modeling is a way to describe behaviors that are not deterministic by na-
ture, or to abstract a behavior that is simply too complex to model explicitly. So it is
typically very useful in a SoS context.
This paper proposes extension of the UPDM/SysML language, depicted in Fig.~\ref{fig:stereotypes},
based on a set of attribute stereotypes that can be applied to any block attribute.

Stochastic behavior is classically represented by the concept of random variable. The
probability distribution of the variable is interpreted as the probability that the varia-
ble takes up a certain value when it is observed. To follow this classical approach,
given a random variable R that takes values over the reals and a real variable V, the
assignment ``V = R\_observe()'' can be seen as an observation of the random variable,
provided that ``R\_observe()'' function is automatically or manually defined to generate
new random values.

A simple example to illustrate this could be the statement ``real initialForce =
force\_observe()'', where all such assignment statements are interpreted as distinct
observations of the variable force. For random timing delays, one would have to in-
stantiate a random real variable ``delay'' and to use it as ``tm(delay\_observe())'', where
``tm()'' is a built in function in Rhapsody SysML to access simulation time.
Based on these remarks, following is the list of the proposed stochastic stereotypes
(inside the DANSE profile):

\begin{itemize}
\item UniformRandomReal $\rightarrow$ automatic addition of ``RhpReal observe()'' at code generation for generating a real based on uniform distribution
  \begin{itemize}
  \item min:String
  \item max:String
  \end{itemize}
\item UniformRandomInteger $\rightarrow$ automatic addition of ``RhpInteger observe()'' at code
generation for generating an integer based on uniform distribution
  \begin{itemize}
  \item min:String
  \item max:String
  \end{itemize}
\item NormalRandomReal $\rightarrow$ automatic addition of ``RhpReal observe()'' at code generation for generating a real based on normal distribution
  \begin{itemize}
  \item mean:String
  \item standardDeviation:String
\end{itemize}  
\item NormalRandomInteger $\rightarrow$ automatic addition of ``RhpInteger observe()'' at code
  generation for generating an integer based on normal distribution
  \begin{itemize}
  \item mean:String
  \item standardDeviation:String
  \end{itemize}
\item CustomRandomReal $\rightarrow$ manual addition of ``RhpReal observe()'' to generate the
  random numbers (based on any kind of distribution)
  \begin{itemize}
  \item customObserveFunction:String
  \item CustomRandomInteger $\rightarrow$ manual addition of ``RhpInteger observe()'' to generate
    the random numbers (based on any kind of distribution)
  \item customObserveFunction:String
  \end{itemize}
\end{itemize}

The reason for declaring all the attributes above as ``String'' is to be able to specify expressions and not only numerical values (e.g., the mean value of the normal distribution of the city traffic is probably not fixed but dependent on the time of the day and the day of the week). In addition to this, it would be worth considering the creation of an API allowing the overwriting of these expressions at any simulation time instant.

\section{The GCSL language and the modeling of SoS contracts}
The objective of the Goal Contract Specification Language (GCSL) is to provide a language specifically designed to specify goal requirements for systems of systems. The GCSL syntax is a combination of the Object Constraint Language (OCL) \cite{OCL} and the contract patterns à la SPEEDS \cite{project-SPEEDS}. The SPEEDS contract specification patterns are introduced in the SPEEDS Deliverable D.2.5.4 ``Contract Specification Language (CSL)'' and used to give a high-level specification of real-timed components. They have been introduced to enable the user to reason about event triggering that is equivalently replaced in this document by constraint satisfaction. The constraints handled by these patterns are about the state of a SoS. We use OCL to specify these state constraints. The Object Constraint Language is a well-established standard in software design. The language is sufficiently powerful to describe precisely a state of a SoS. In this document, we will only consider a subset of the OCL language. In particular, we mainly focus on the Collection type without considering all its refinements (Set, Ordered Sets, ...) and the most common predicates about Collections (forAll(x$|$...), exists(y$|$...), select(...), ...). The relevant part of the language considered is the expression subset of the language, i.e. Boolean, Collection or arithmetic expressions.

The GCSL patterns are designed to specify the behavior of each component instance by totally abstracting its environment without quantification. It is not possible to specify a contract about the interaction between two anonymous components. By anonymous, we mean that no particular instance is explicitly referenced by the component identifier. It is an important limitation for DANSE: to overcome it, we introduce a way to express quantifications that overlap a contract pattern. We use the OCL predicates forAll() and exists() over OCL Collections to introduce the capability to quantify over the instances of any component type of an SoS. For additional details the reader is referred to \cite{AiSoS2013Legay}.

\section{Statistical model checking tool-chain}

\begin{figure}[h]
  \centering
  \includegraphics[width=0.9\linewidth]{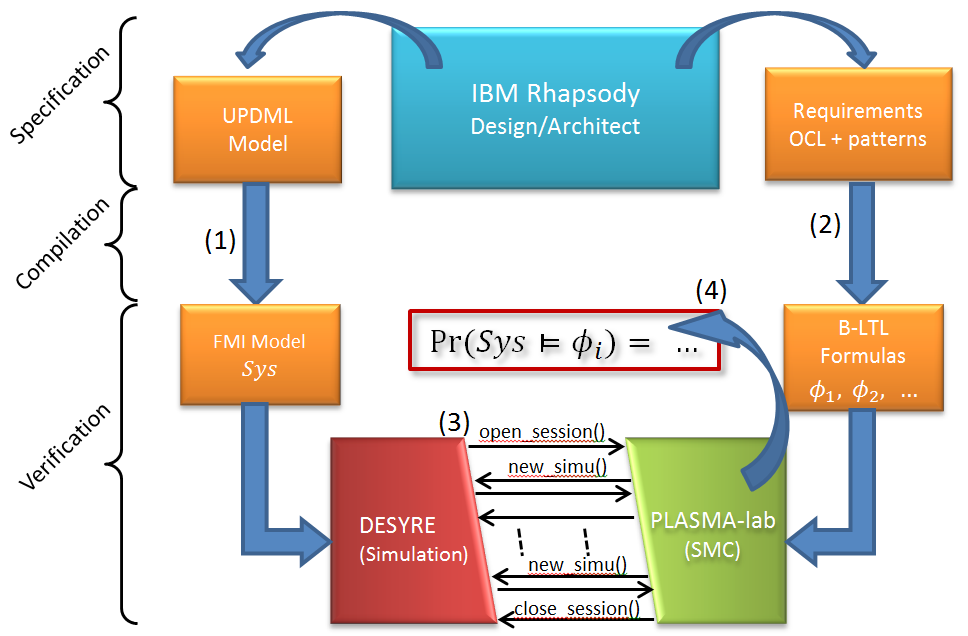}
  \caption{Overview of SMC process in DANSE}
  \label{fig:toolchain}
\end{figure}

Fig.~\ref{fig:toolchain} shows the statistical model checking tool-chain. The SoS is modeled in UPDM/SysML and annotated with GCSL contracts. The specification follows two paths: the GCSL contracts are translated to the B-LTL language accepted by PLASMA and enriched with OCL patterns. The structure of the modeled follows a different technological path to be imported into the DESYRE simulator. The model structure is annotated with the referenced component's implementations.

\subsection{PLASMA: a statistical model checking tool}
PLASMA \cite{Plasma-Lab}, developed at INRIA \cite{INRIA}, adopts a modular architecture to facilitate the extension of its features. Models can currently be specified using the PRISM \cite{PRISM} reactive modules syntax, timed automata and biologic systems, but the implementation of other modeling formalisms, such as procedural programming languages such as C and Java, is already planned.

\begin{figure}[h]
  \centering
  \includegraphics[width=0.7\linewidth]{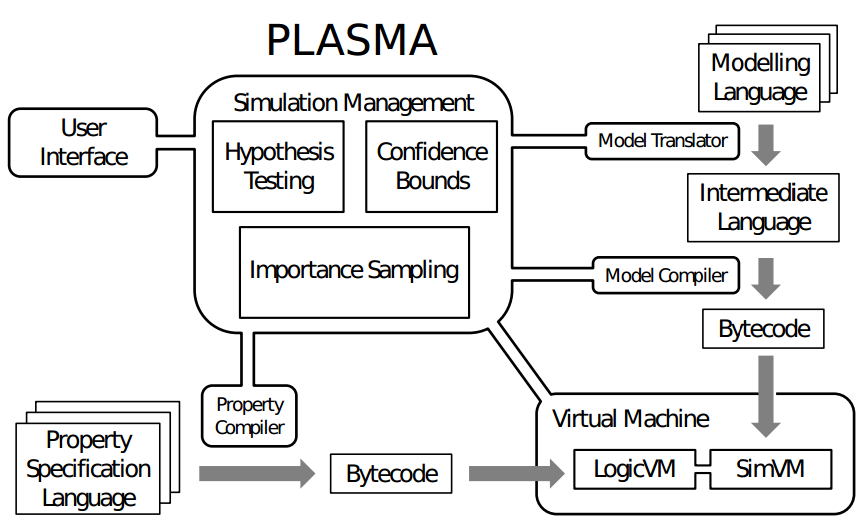}
  \caption{The architecture of PLASMA}
  \label{fig:plasma}
\end{figure}

The input specification is translated into a common intermediate language based on elements (referred to as simple commands because they have no explicit synchronization and no choice of actions) having the structure (guard, rate, actions), where guard, rate and actions are functions over the current state (constants, variables, clocks) of the system. The intermediate language thus expresses the semantics of a system which advances by discrete events: the guard enables the command, the rate resolves non-determinism between enabled commands (and controls the delay in continuous time systems) and the actions update the state of the system. Once the model is represented in the intermediate language it is compiled into an executable form (the model program). PLASMA uses its own in-built compiler to create bytecode for execution on its own stack-based virtual machine (VM) that comprises a logic VM (LogicVM in Fig.~\ref{fig:plasma}) and a simulation VM (SimVM in Fig.~\ref{fig:plasma}). PLASMA’s bytecode instructions constitute a domain-specific, low level, platform-independent language designed for efficient statistical model checking. This language contains standard low level instructions, such as push, pop, add, sub, mul, div, etc., as well as non-standard instructions to construct efficient model checking algorithms. The VM is implemented in a high level procedural programming language (currently Java, but the code uses no features that cannot easily be adapted to other languages) and is efficient because it is optimized for its domain of application: high level instructions are efficient sub-parts of model-checking algorithms and all instructions are optimized with respect to the hardware level. The compiler and VM are also sufficiently compact to allow PLASMA to be implemented as a browser application, a distributed component or in an embedded system.

PLASMA verifies properties specified in bounded temporal logic. Such properties are compiled into bytecode programs (property programs) and then executed on the logic VM. Our current focus is discrete time, however continuous time and other logics may be easily facilitated by implementing additional logic parser-compilers. Overall control of the verification process is maintained by the simulation management kernel (SMK) according to the options specified by the user. In general, the property program executes the model program until it has seen sufficient steps to decide a result and the SMK executes the property program until it has sufficient results to return an answer to the user. This way, simulation traces contain the minimum number of states necessary to decide the property and the minimum number of simulations are generated. The logic accepts arbitrarily nested path formulae, however formulae that are not nested are particularly memory efficient: by employing a multivalued logic (true, false, undecided) PLASMA need only store the current state of the system. Nested formulae are also handled efficiently. In general, PLASMA stores only a subset of the full trace, having length equal to the maximum sum of the time bounds of any nested formulae.

\subsection{DESYRE simulation environment}
DESYRE \cite{SIES2012Ferrari,ERTS2012DAngelo,ERTS2012Ferrari} is a System-Level Simulation Framework developed by ALES S.r.l. \cite{ALES}. It is a SystemC \cite{accelera} based simulation environment used to support the design and the verification of Large-Scale Real-Time Distributed Embedded Systems. The tool provides capabilities to perform Design Space Exploration and System Verification by simulation. It has been successfully employed in several EU \& DARPA Projects such as: SPEEDS \cite{project-SPEEDS}, METAII \cite{ERTS2012Ferrari,DARPAMETA}, SPRINT \cite{SPRINT}, MBAT \cite{MBAT} and in several industrial projects \cite{SIES2012Ferrari, ERTS2012DAngelo}.

DESYRE is a component based simulator where system models are described as the composition of a set of modules. Hierarchical modules are supported. Modules are arranged in libraries. A library represents a group of models that are logically related (e.g., the IEEE802.3 library contains the models of the physical and MAC layer defined by the IEEE standard).

A module in DESYRE is fully identified by its interface, which specifies the set of parameters and ports (interaction points) of the module, and by a set of implementations, that represent the behavior of the module at different levels of abstraction. DESYRE modules have an associated descriptor that is necessary to enable the instantiation and the configuration of the module as a component into a system model definition. 

A DESYRE system model can be specified either with SPIRIT IPXACT \cite{IP-XACT} XML standard files or with SLang (System Language for DESYRE) XMI files. This way the user has the capability to specify its own system model and simulate several configurations, without the need to recompile the entire model. 

\begin{figure}[h]
  \centering
  \includegraphics[width=0.7\linewidth]{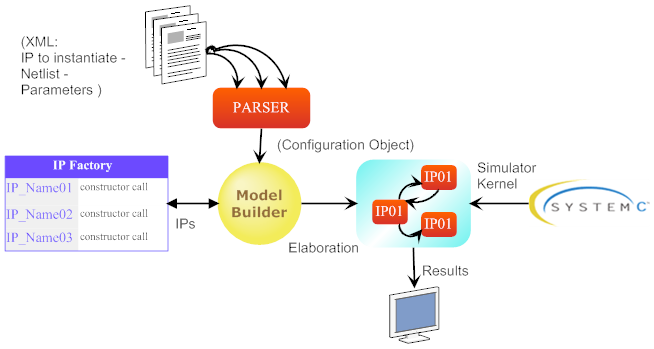}
  \caption{DESYRE model building and simulation}
  \label{fig:desyre}
\end{figure}

The architecture of the DESYRE framework is shown in Fig.~\ref{fig:desyre}. To run the simulation, the system descriptor files are provided to the DESYRE simulation kernel. The kernel parses the files and invokes the DESYRE model builder that instantiates the required modules, interconnects them and configure their parameters. As soon as the model has been completely instantiated, the SystemC simulation kernel is invoked and the simulation starts. During the simulation, the simulator collects logs and traces that can be used later on to create diagrams or analyze system performances.

The main features of the simulator are:
\begin{itemize}
\item Simulation of heterogeneous components (capability to import models from other
tools or defined with other languages, such as: MATLAB/Simulink \cite{mathworks}, Modelica \cite{modelica}, Rhapsody/SysML \cite{rhapsody}, FMI \cite{FMI});
\item Hardware and software in the loop simulation (capability to simulate a system
model connected to real physical devices and/or interacting with some real target
software components);
\item Fault Injection Capabilities to support fault tolerance analysis;
\item Tracing mechanisms;
\item Monitor synthesis generation (capability to generate executable monitors starting
from contract specifications);
\item Support for debugging of code executed on the simulated system;
\item Support for distributed simulation using the HLA standard \cite{xiaoxia2003mpii}.
\end{itemize}

\subsection{Functional mock-up interface (FMI)}
The Functional Mock-up Interface (FMI) \cite{FMI} is a tool- and language-independent
open standard to support heterogeneous models composition and simulation with
different levels of detail. FMI was developed in the MODELISAR ITEA2 \cite{ITEA} project
(2010-2012) \cite{modelisar} and its specification was initiated by Daimler AG.
To support the FMI standard a modeling environment must be able to generate C-
Code from a dynamic system model. The code can be used by other modeling and
simulation environments to perform simulation and analysis. An executable model
that implements the FMI interface is called Functional Mock-up Unit (FMU). FMI
models can either be self-integrating (FMI for Co-Simulation) or require the simula-
tion environment to perform numerical integration (FMI for Model Exchange). The
FMU can be imported into a simulation environment that uses the FMI functions to
create model instances, interconnect them to other models and execute them. FMI is a
component-oriented standard, i.e. no system representation that specifies how compo-
nents are interconnected and interact is available.

\subsection{FMI support in DESYRE}
\begin{figure}[h]
  \centering
  \includegraphics[width=0.6\linewidth]{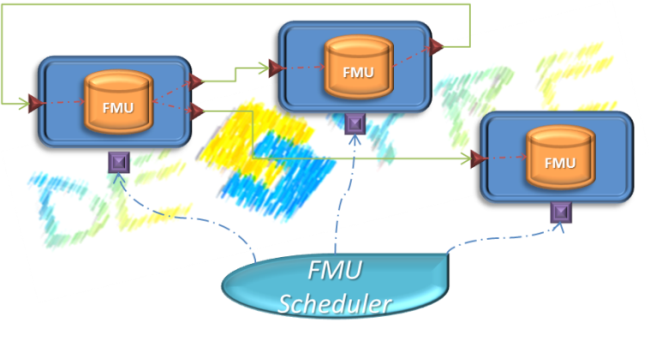}
  \caption{FMI support in DESYRE, FMU interconnection and architecture}
  \label{fig:FMI}
\end{figure}

DESYRE supports the import and simulation of functional mockup units (Fig.~\ref{fig:FMI}) for
both Model Exchange and Co-Simulation version of the standard. In the context of
the DANSE EU project, the FMI standard version 1.0 for Model Exchange will be
used.
DESYRE API provides a service \linebreak[4] (``des\_importFMU(String fmuURI)'') that allows
the user to import an FMU into the DESYRE workspace and automatically generates
the DESYRE module descriptor necessary to enable the DESYRE model builder to
instantiate and interconnect the new module into a system model.
Imported FMUs are wrapped with an FMI Wrapper module that allows the
DESYRE simulator to set and get FMU inputs, outputs and parameters and to correct-
ly solve the FMU model equations. FMU I/O are redirected on a DESYRE module
I/O ports.

To interconnect two or more FMUs it is enough to connect their I/O ports using the
DESYRE XML or SLang specification language.

When an FMU model is instantiated, the DESYRE simulator automatically regis-
ters it to the DESYRE FMI Scheduler. The DESYRE FMI scheduler is an FMI 1.0
compliant scheduler that requires the FMUs execution according to the FMI standard
protocol rules.

\subsection{PLASMA/DESYRE integration and SMC for SoS}
\begin{figure}[h]
  \centering
  \includegraphics[width=1\linewidth]{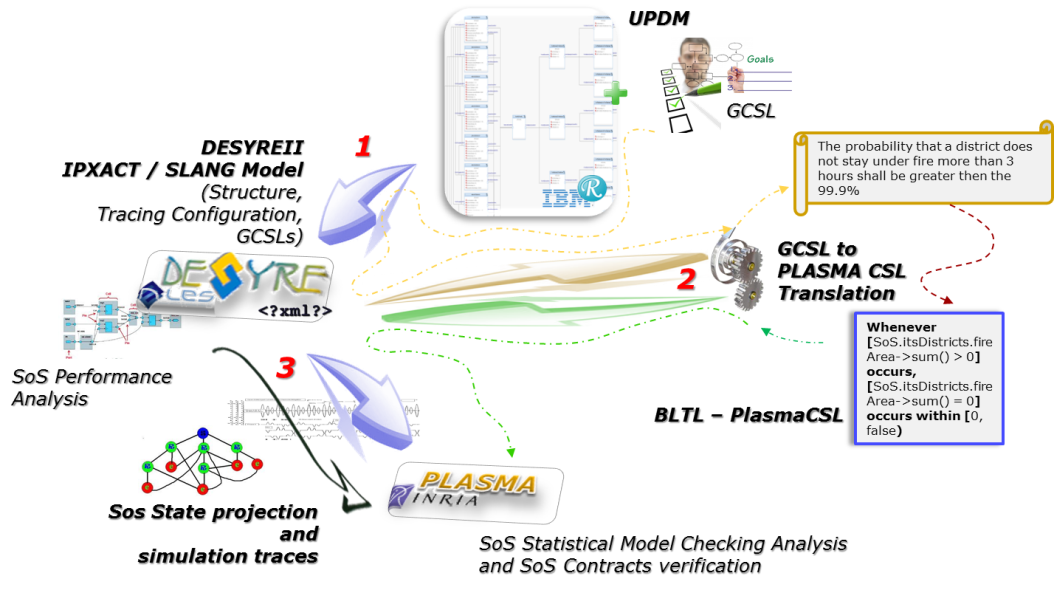}
  \caption{DESYRE/PLASMA Statistical Model Checking flow}
  \label{fig:flow}
\end{figure}

Fig.~\ref{fig:flow}. summarizes the workflow to perform a statistical model checking analysis of a
SoS model. The SoS model is specified in IBM Rhapsody using the UPDM language.
Using the DANSE profile for UPDM the designer is able to annotate the model with a
set of goals, contracts and properties that he would like to verify using the simulation
and analysis tools. Such annotations are expressed using the Goal Contract Specifica-
tion Language (GCSL) introduced earlier. Once the SoS model is complete, a model
transformation flow will translate the UPDM model into a DESYRE SLang model
and will extract the GCSL expressions from the model annotations. GCSL properties
are then translated into B-LTL expressions that can be understood by PLASMA
(PlasmaCSL).

Once the model transformation is complete, the designer can carry out the analysis of
the SoS model. The DESYRE/PLASMA statistical model checking user interface
allows the designer to configure and invoke the analysis on the SoS and to retrieve the
analysis results at the end of the elaboration.

\subsection{DESYRE/PLASMA SMC configuration}
\begin{figure}[h]
  \centering
  \includegraphics[width=1\linewidth]{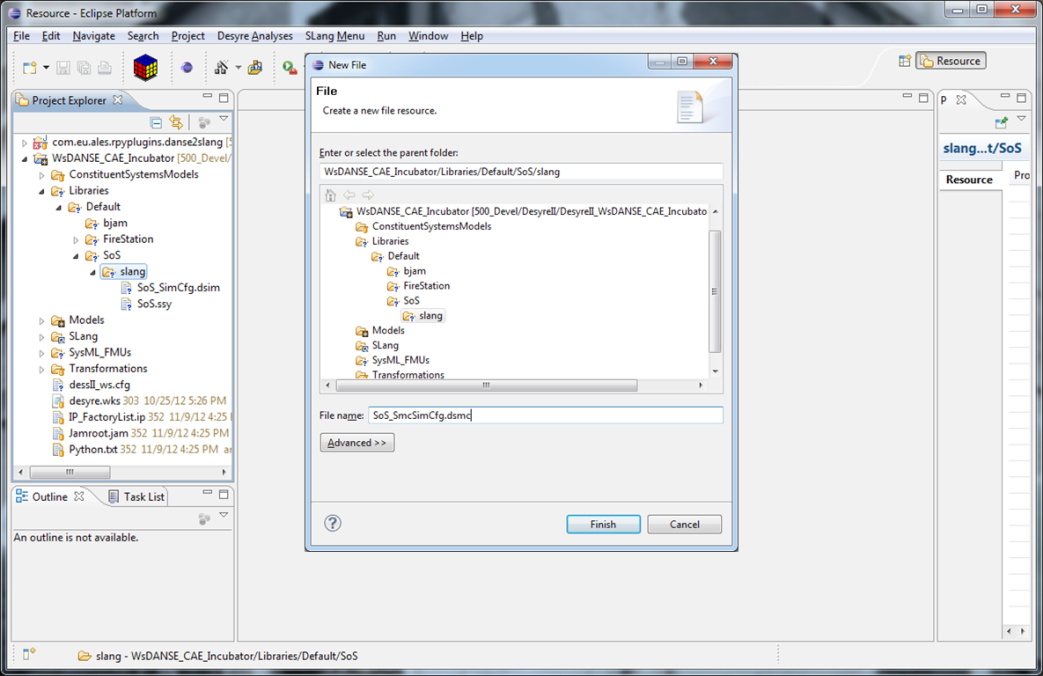}
  \caption{Create a DESYRE/PLASMA Statistical Model Checking Session}
  \label{fig:session}
\end{figure}

To run the Statistical Model Checking of the SoS model, the user is required to create
and configure the DESYRE/PLASMA statistical model checking session. To create a
statistical model checking session it is necessary to create a new ``.dsm'' file into the
DESYRE workspace as shown in Fig.~\ref{fig:session}. By opening the file the user will be able to
select the SoS model to analyze and specify the kind of statistical analysis technique
to use and the set of properties and contracts to verify.

\begin{figure}[h]
  \centering
  \includegraphics[width=1\linewidth]{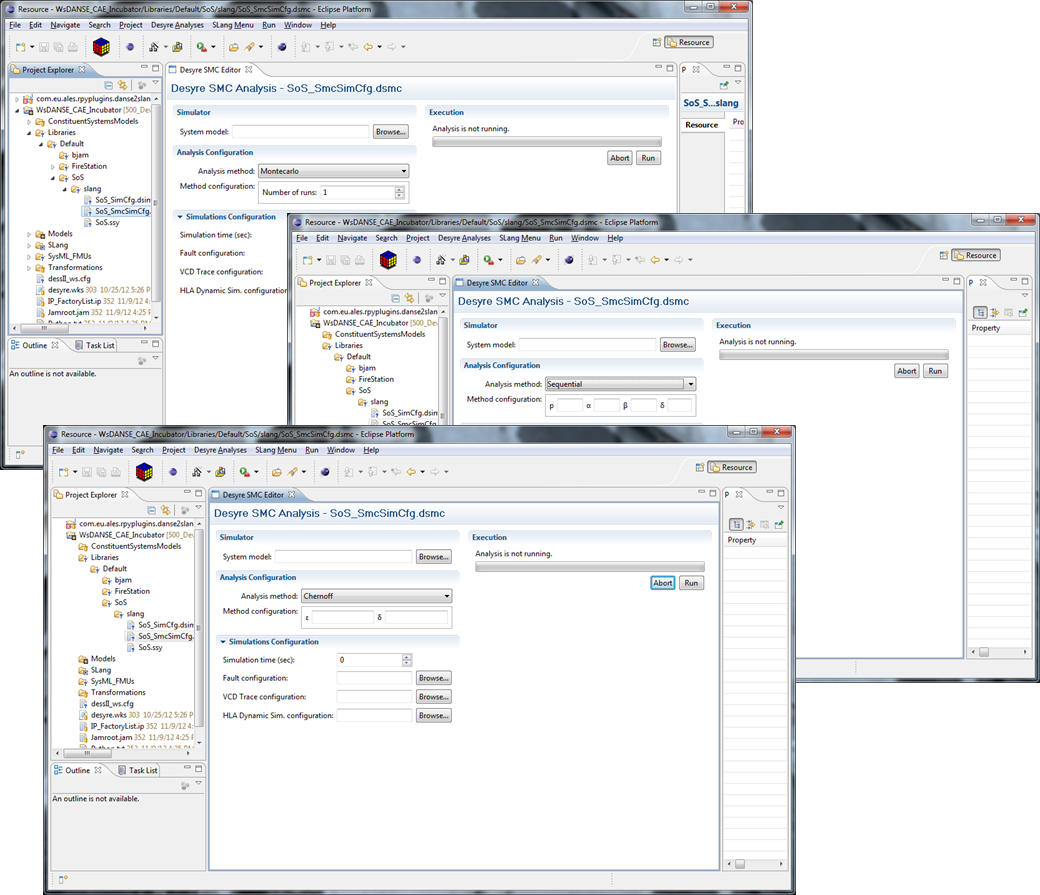}
  \caption{DESYRE/PLASMA Statistical Model Checking session configuration}
  \label{fig:configuration}
\end{figure}

Three different analyses techniques are available: Montecarlo simulations, error bounded probability estimation with Chernoff-Hoeffding bound and hypothesis testing with the sequential ratio probability test. The set of input parameters that the user has to specify may change depending on the selected analysis technique (Fig.~\ref{fig:configuration}). Pressing the ``Run'' button starts the SMC analysis session. 

\begin{figure}[p]
  \centering
  \includegraphics[width=0.9\linewidth]{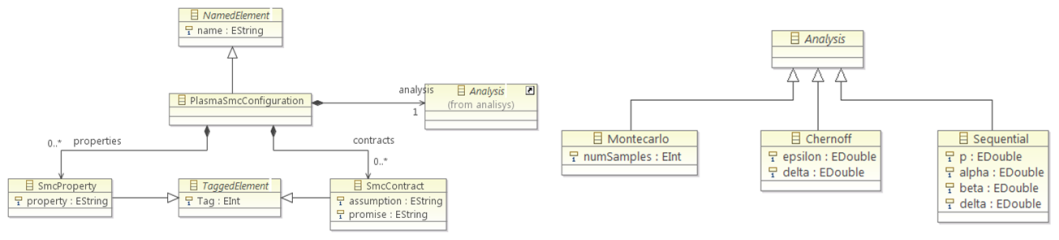}
  \caption{PLASMA SMC Configuration Object}
  \label{fig:object}
\end{figure}

The SMC analysis configuration contains information on the selected statistical analysis technique and the set of properties and contracts to verify, as showed in Fig.~\ref{fig:object}.

\begin{figure}[p]
  \centering
  \includegraphics[width=0.5\linewidth]{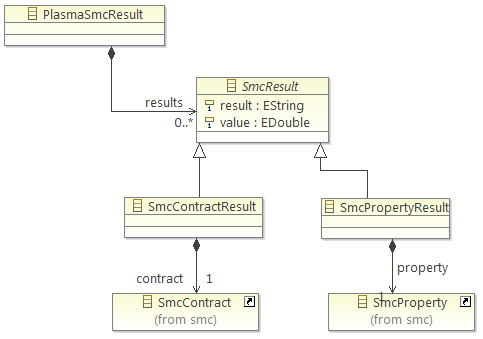}
  \caption{PLASMA Results Object}
  \label{fig:results}
\end{figure}

At the end of the analysis session PLASMA returns the analyses results in a structured object, associating a result element with each property and contract specified in the PLASMA configuration object as represented in Fig.~\ref{fig:results}.

\subsection{DESYRE/PLASMA SoS Model State}
\begin{figure}[p]
  \centering
  \includegraphics[width=0.7\linewidth]{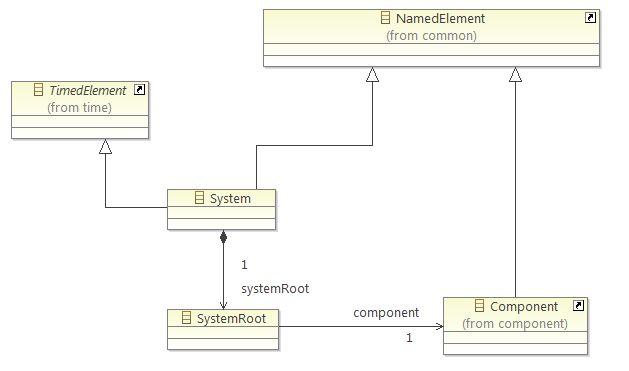}
  \caption{DESYRE Simulation State Projection - System}
  \label{fig:system}
\end{figure}

\begin{figure}[h]
  \centering
  \includegraphics[width=0.7\linewidth]{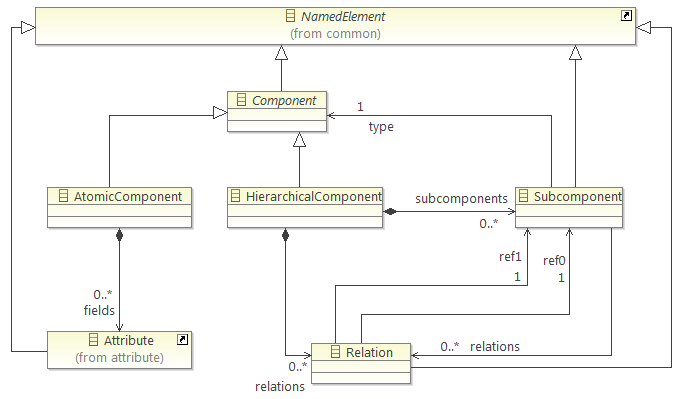}
  \caption{DESYRE Simulation State Projection - Component}
  \label{fig:component}
\end{figure}

In order to perform its computation and evaluate the specified properties, PLASMA needs to refer to the actual structure of the SoS model and the value of its constituent systems inputs, outputs and attributes for each simulation step. This has been
achieved by defining a simulation state class of objects that represent the projection of the simulated model state on a simplified structure that can be easily elaborated by PLASMA (Fig.~\ref{fig:system} and Fig.~\ref{fig:component}).

The system state projection is represented by the ``System'' (Fig.~\ref{fig:system}). The ``System'' contains a ``System Root'' element that points to the SoS top-level component. The top-level is usually a hierarchical component.

A “Component'' can either be atomic or hierarchical. An ``Atomic Component'' represents the behavioral elements of the SoS (the Constituent Systems) and contains a set of fields (``Attribute''). An ``Attribute'' has a name, a type and an actual value. A ``Hierarchical Component'' contains a set of ``Subcomponents'' and a set of ``Relations''. A ``Subcomponents'' has a name and an associated type (``Component''). A ``Relation'' expresses the existence of an interaction between two ``Subcomponents'' and has an associated name and two references to two different ``Subcomponents''. Each ``Sub-component'' has a list of references to the ``Relation'' elements that refer to him.

\subsection{Closed and open systems}
In a closed system, all the variables and types are known a priori, hence the state vector has constant width throughout the simulation and the requirements of the interface are straightforward. E.g., in a model of an emergency system where the number of ambulances available is given by the value of an integer variable, , adding or removing an ambulance is simply a matter of changing . In a system with a fixed number of ambulances, the ambulances might also have properties, such as an amount of fuel and number of personnel, that can be referred to explicitly, e.g., a.fuel. This is the kind of model we have been working with so far and is the easiest class to accommodate in an initial integration of PLASMA and DESYRE.
In an open system, in particular in a dynamically configured system of systems (SoS), variables may appear and disappear during the course of the simulation as new complex entities enter and leave the system. E.g., in an emergency system where vehicles with individual characteristics may be added or removed, each change adds or removes a set of local variables that describe the characteristics of the vehicle. It is not clear if it is useful to allow property formulae to refer to the local variables of subsystems that do not yet exist or that cannot be guaranteed to exist in the future, but a plausible approach would be to add quantifiers to the logic that reason on classes of variables. E.g., $G_{\leq t} \forall a \in Ambulances , a.fuel > 0$ meaning that no ambulance will run out of fuel in ``t'' time units.

To accommodate open systems is clearly trickier, but we know that this will be an eventual requirement so it will be an advantage to comment some of the issues now. In particular, when the number and position of variables in the state vector data structure change during the course of a simulation of an open system, PLASMA will be required to search for the variables it requires. To avoid unnecessary computation we might include flags that indicate whether the system is open or closed and, if open, when the variables change (the latter may be sufficient to achieve both objectives). In this way, PLASMA need only search when variables actually change. Similarly, when the intermediate class requests a block of states from DESYRE, the size of the block supplied may be determined by providing sets of states with consistent variables.

\subsection{Conclusion}
In this paper we have presented a statistical model checking framework for systems of systems, developed with support from the DANSE EU FP7 project. The framework focuses on performance constraints and non-critical structural and temporal constraints. We have shown how the extensible UPDM/SysML language for systems of systems can be successfully extended to represent stochastic behavior and complex structural and temporal contract specifications. We have presented an innovative tool-chain that integrates two successful technologies, namely PLASMA, developed at INRIA, and DESYRE, developed at ALES S.r.l., to effectively provide scalable performance verification techniques for systems of system.

\subsection{Acknowledgement}
The authors would like to acknowledge the support of the DANSE EU project (grant agreement no. 287716).

\bibliographystyle{eptcs}
\bibliography{biblio}
\end{document}